\documentclass[sigconf]{acmart}

\renewcommand\footnotetextcopyrightpermission[1]{}

\usepackage{amsmath}
\usepackage{tcolorbox}
\usepackage{tikz}
\usetikzlibrary{shapes,positioning}

\AtBeginDocument{%
  \providecommand\BibTeX{{%
    \normalfont B\kern-0.5em{\scshape i\kern-0.25em b}\kern-0.8em\TeX}}}

\setcopyright{acmcopyright}
\copyrightyear{2022}
\acmYear{2022}
\acmDOI{XXXXXXX.XXXXXXX}

\acmConference[KDD]{KDD}{
  XXX}{XXX}

\acmPrice{XXX}
\acmISBN{XXX}

\acmSubmissionID{XXX}

\begin{document}

\title{Valid and Unobtrusive Measurement of Returns to Advertising through Asymmetric Budget Split}

\author{Johannes Hermle}
\email{jhermle@linkedin.com}
\affiliation{%
  \institution{LinkedIn Corporation}
  \country{}
}

\author{Giorgio Martini}
\email{gmartini@linkedin.com}
\affiliation{%
  \institution{LinkedIn Corporation}
   \country{}
}

\begin{abstract}

Ad platforms require reliable measurement of advertising returns: what increase in performance (such as clicks or conversions) can an advertiser expect in return for additional budget on the platform? Even from the perspective of the platform, accurately measuring advertising returns is hard. Selection and omitted variable biases make estimates from observational methods unreliable, and straightforward experimentation is often costly or infeasible. We introduce \textit{Asymmetric Budget Split}, a novel methodology for valid measurement of ad returns from the perspective of the platform. Asymmetric budget split creates small asymmetries in ad budget allocation across comparable partitions of the platform's userbase. By observing performance of the same ad at different budget levels while holding all other factors constant, the platform can obtain a valid measure of ad returns. The methodology is unobtrusive and cost-effective in that it does not require holdout groups or sacrifices in ad or marketplace performance. We discuss a successful deployment of asymmetric budget split to LinkedIn's Jobs Marketplace, and ad marketplace where it is used to measure returns from promotion budgets in terms of incremental job applicants. We outline operational considerations for practitioners and discuss further use cases such as budget-aware performance forecasting. 
\end{abstract}


\maketitle

\pagestyle{fancy}
\fancyhf{}
\rhead{\thepage}
\lhead{}
\cfoot{}

\section{Introduction}

\subsection{Motivation} 

Online advertising constitutes a substantial source of revenue for modern technology companies and is expected to grow \cite{evans2009online, parssinen2018blockchain}. Ad platforms rely on complex value delivery systems \cite{goldfarb2019digital, varian2010computer} to optimize returns to advertisers. In order to assess the returns delivered to advertisers through these systems, ad platforms require reliable and credible measurement: what is the additional return for an advertiser for each each additional dollar in budget on the platform?

Reliable measures of advertising returns are of primary importance for ad platforms. First, they are needed for internal performance management. Second, they are needed to evaluate the impact of improvements to their value delivery systems. Third, advertisers demand information about expected costs and returns of investing on the platform. Without reliable measures of returns to spend, ad platforms are not able to provide confidence to advertisers about their investment.

Accurately quantifying returns to advertising from the platform's perspective, however, is hard: an ad platform's value delivery systems typically rely on complex online auctions \cite{ockenfels2006online} and involve manifold sub-components that govern bidding strategies \cite{zhang2014optimal} and budget pacing \cite{xu2015smart,agarwal2014budget}. Even with full knowledge of these systems, their complexity makes it challenging to reliably quantify returns to advertising. 
Proxy metrics commonly employed by business analysts such as ``cost-per-click'' are problematic as they are not necessarily incremental \cite{hu2016incentive}: a certain user action desired by the advertiser may occur regardless of whether the user was shown an ad or not \cite{lewis2011here,farahat2012effective}.

Despite a growing body of research, current approaches to measuring advertising returns have been highlighted as inefficient \cite{gordon2021inefficiencies}. Even though rich and granular data are available to most ad platforms, observational methods common in the industry often fail to produce credible estimates \cite{gordon2019comparison}. Specifically, observational approaches often suffer from selection and omitted variable bias \cite{lewis2015unfavorable}. 
Hence, correlating ad budget with performance does not provide a satisfactory solution; ads with higher budgets may see higher performance potentially due to their higher user relevance and not due to budget.

A direct experiment that addresses this identification challenge would be to randomly perturb budgets across ads. By introducing an exogenous source of variation, the platform could causally estimate the expected increase in performance per additional unit of budget. However, such a direct experiment is infeasible or at least impractical. The platform cannot unilaterally alter an advertiser's budget as this would either harm the advertiser's performance or increase their costs. 

\subsection{Asymmetric budget split}

To overcome these challenges, we introduce \textit{asymmetric budget split}, a novel method that allows ad platforms to obtain a valid measure of advertising returns.
Asymmetric budget split offers an unobtrusive method to introduce exogenous variation in budgets. Instead of changing the ad's \textit{total} budget, the method creates two copies of every ad, with the total budget split \textit{asymmetrically} (i.e. unequally) among them (but otherwise identical). The two ad copies participate in isolated \textit{sub-markets} of the ad platform, which are otherwise comparable. The analyst can thus observe the ad's performance at different budget levels while holding other factors constant. By relating the change in ad performance to the (exogenous) difference in budget, the analyst can causally estimate the marginal return of additional budget.

The formulation and implementation of sub-markets used by asymmetric budget split is shared with the regular budget-split design \cite{liu2020trustworthy}. 
Neither the (symmetric) budget-split design nor asymmetric budget split restrict the matching between users and ads: all ads are potentially exposed to all users. This is because each ad is split into copies, one per sub-market. The main implementation difference between the two methodologies is that under asymmetric budget split the budget of each ad copy is not fixed to be in proportion to the sub-market size. As such, the additional engineering cost of deploying asymmetric budget split to platforms that already have the infrastructure needed to run budget-split tests is minimal.

Importantly, the budget-split design \cite{liu2020trustworthy} and asymmetric budget split answer different classes of questions. Budget-split design was developed to run valid user-level experiments, which are otherwise subject to interference bias due to users in different experiment arms sharing budgets on the other side of the marketplace. Asymmetric budget split, on the other hand, is used to measure ads returns. Both designs are compatible and can be employed by the platform at the same time.

\subsection{Our contribution}

Asymmetric budget split contributes to the body of work on causal measurement of ad returns \cite{gordon2021inefficiencies}. Recent advances in this literature have been made in particular through experimental approaches \cite{goldfarb2011online, johnson2017ghost, lewis2014online}. Our methodology is distinct in three ways.

First, while asymmetric budget split is inherently `experimental' in that it relies on randomized allocation of budgets across sub-markets, it does not incur the measurement costs often associated with  experiments \cite{lewis2015unfavorable}. Such costs can be substantial particularly in designs where users in a control group are held out from being displayed an ad \cite{hoban2015effects, sahni2015effect, johnson2017less}. In such cases, the ad platform forgoes a monetization opportunity. Asymmetric budget split does not directly affect the pattern or rate at with which ads are impressed, and there are no holdout or control groups.

Second, asymmetric budget split differs from most existing experimental approaches in terms of where randomized variation is introduced. Most existing experimental designs introduce randomized variation in ad exposure across users or sessions. This allows the analyst to quantify the causal effect of showing an ad on user behavior. Asymmetric budget split instead introduces random variation in budget. This allows a holistic analysis which directly measures the causal relationship between budget and ad performance.

Third, while we focus the discussion on measurement of ads returns, the asymmetric budget split method is general. It is directly suited to all types of online advertising, from keyword ads to display ads to sponsored listings, where advertisers regulate their participation by setting budgets (instead of setting bids directly).
Beyond the online ads settings, it can be used in any environment where a divisible and exhaustible quantity (``budget'') influences outcomes, and where the marginal impact of ``budget'' on outcomes is of interest to the analyst. Budgets do not have to be monetary: examples include resource constraints such as inventories; quotas for push notifications or in-app promos; and impression limits for items.

The remainder of this paper is structured as follows. In Section \ref{sec:causal_measurement}, we first describe the technical setup of asymmetric budget split and econometric details of ads return measurement. In Section \ref{sec:architecture}, we discuss the data and system architecture. In Section \ref{sec:application}, we review an application to LinkedIn's Jobs Marketplace where asymmetric budget split was successfully deployed in production and is used to estimate ads returns on the platform. In Section \ref{sec:considerations}, we discuss operational considerations for ad platforms when implementing asymmetric budget split as well as further use cases before we conclude in Section \ref{sec:conclusion}.

\section{Methodology}\label{sec:causal_measurement}
\subsection{Identification challenge and solution}
Consider an ad $i$ with budget $B_i$ to be spent on the platform.
For ease of exposition we use the term `ad' throughout, but our methodology applies unchanged to any form of sponsored or promoted listings, including but not limited to search results on search engines, 
sponsored items on e-commerce marketplaces, or
job postings on online job boards.

Let $Y_i$ be the performance of ad $i$, measured as the number of clicks, conversions, or any other outcome desired by the advertiser. $Y_i$ is a function of budget $B_i$ and ad features $X_i$ (such as targeting criteria and ad format). We are interested in estimating advertising returns: the marginal increase in performance $Y_i$ due to one unit of additional budget $B_i$, which we label \textit{incrementality}.\footnote{In the literature and practice of marketing and online advertising, `incrementality' often refers to the causal impact of an ad exposure on user behavior, such as purchasing a product \cite{lewis2018incrementality}. Our definition of incrementality is expressed in terms of an additional unit of budget, not of an additional ad exposure.}

Regression models that use observational data to estimate the relationship between $Y_i$ and $B_i$ are ill-suited to provide a valid estimate of incrementality, because budget $B_i$ is \textit{non-random} or endogenous. Correlational relationships between $Y_i$ and $B_i$ obtainable from observational data often suffer from omitted variable bias, which can create substantial bias in the estimate \cite{lewis2015unfavorable}.
For instance, an advertiser who expects her ad to perform well may set a higher budget, leading to a spurious positive correlation between performance and budget. This results in an upward bias in incrementality estimates from observational data.
Perverse effects that introduce a downward bias can also occur. For example, an advertiser who aims to reach a certain performance level may set a higher budget precisely when she expects low performance per unit of spend. This introduces a spurious negative correlation between budget and performance, which in extreme cases can lead to \textit{negative} incrementality estimates even when the true causal impact of additional budget is positive. 
In sum, as the process of budget setting is typically unknown to the analyst and depends on unobservable factors of the ad or the advertiser, correlational estimates that rely on non-random variation in budget can have substantial bias. Credible identification of incrementality requires (quasi-)random variation in $B_i$. 

To solve this identification challenge, we introduce \textit{asymmetric budget split} as a novel method for valid incrementality measurement on ad platforms. Asymmetric budget split does not change an ad's total budget to be spent on the platform's ad market. Instead, it introduces random perturbations in the fraction of an ad's budget allocated to distinct \textit{sub}-markets of the ad platform. The sub-markets consist of randomly selected, equal-sized and disjoint subsets of the overall user base of the platform. Hence, they are balanced in terms of their characteristics. 

This procedure allows the analyst to study performance for the same ad 
at different budget levels. By relating the differences in ad performance to the corresponding differences in budget, the analyst obtains a valid measure of incrementality. 

The following sections describe the implementation of asymmetric budget split before outlining the econometric details of incrementality estimation.

\subsection{Asymmetric budget split}\label{ssec:abs_methodology}

\begin{figure}[h]
\begin{tikzpicture}
 
\node[draw, rectangle, line width=0.5pt, fill=gray!10, rounded rectangle] (U) at (0,1)  {User base $U$};

\node[draw, rectangle, line width=0.5pt, fill=gray!10] (user) at (0,0)  {User $u\in U$};

\node[align=center, draw, line width=0.5pt, fill=gray!10] (U1) at (-1.75,-1) {Sub-market $\mathcal{M}_1$\\ with users $u\in U_1$};
\node[align=center, draw, line width=0.5pt, fill=gray!10] (U2) at (1.75,-1) {Sub-market $\mathcal{M}_2$\\ with users $u\in U_2$};

\draw[-, line width=0.5pt] (U.south) -- (user.north);
\draw[shorten >=0.2cm,shorten <=0.2cm,->, line width=0.5pt] (user) -- (U1.north) node[midway,  above left] {$p = \frac{1}{2}$};
\draw[shorten >=0.2cm,shorten <=0.2cm,->, line width=0.5pt] (user) -- (U2.north) node[midway,  above right] {$p = \frac{1}{2}$};

\node[align=center, draw, line width=0.5pt, text width=24mm, fill=brown!10] (AdHigh) at (-1.75,-4.5) {High-budget\\ ad copy with\\ $B_{i,high} = \alpha\cdot B_i$};
\node[align=center, draw, line width=0.5pt, text width=24mm, fill=brown!10] (AdLow) at (1.75,-4.5) {Low-budget\\ ad copy with \\ $B_{i,low} = (1-\alpha)\cdot B_i$};
\node[draw, rectangle, line width=0.5pt, fill=brown!10] (Ad) at (0,-6)  {Ad $i\in I$ with budget $B_i$};

\node[draw, rectangle, line width=0.5pt, fill=brown!10, rounded rectangle] (Adsample) at (0,-7)  {Set of ads $I$};

\draw[-, line width=0.5pt] (Adsample.north) -- (Ad.south);

\draw[shorten >=0.2cm,shorten <=0.2cm,->, line width=0.5pt] (Ad) -- (AdHigh.south);

\draw[shorten >=0.2cm,shorten <=0.2cm,->, line width=0.5pt] (Ad) -- (AdLow.south);

\draw[shorten >=0.2cm,shorten <=0.2cm,->, line width=0.5pt, dashed] (AdHigh.north) -- (U2.south);
\draw[shorten >=0.2cm,shorten <=0.2cm,->, line width=0.5pt, dashed] (AdLow.north) -- (U1.south);
\draw[shorten >=0.2cm,shorten <=0.2cm,->, line width=0.6pt, densely dotted] (AdHigh.north) -- (U1.south);
\draw[shorten >=0.2cm,shorten <=0.2cm,->, line width=0.6pt, densely dotted] (AdLow.north) -- (U2.south);

\node[line width=0.5pt] (prob1) at (3,-2.2) {$p=\frac{1}{2}$};
\node[line width=0.5pt] (prob2) at (3,-2.8) {$p=\frac{1}{2}$};

\draw[line width=0.6pt, dashed] (2,-2.2) -- (prob1.west);
\draw[line width=0.6pt, densely dotted] (2,-2.8) -- (prob2.west);

\node[align=left] (DescU) at (4.7,-0.5) {Random division\\ of user base\\ into two\\ sub-markets};
\node[align=left] (DescRand) at (4.8,-2.75) {Random allocation\\ of ad copies to\\ the two\\ sub-markets};
\node[align=left] (DescB) at (4.6,-5) {Creation of high-\\ and low-budget\\ ad copies};
\end{tikzpicture}

\caption{Asymmetric budget split schematic}
\label{fig:asym_bs_overview}

\end{figure}
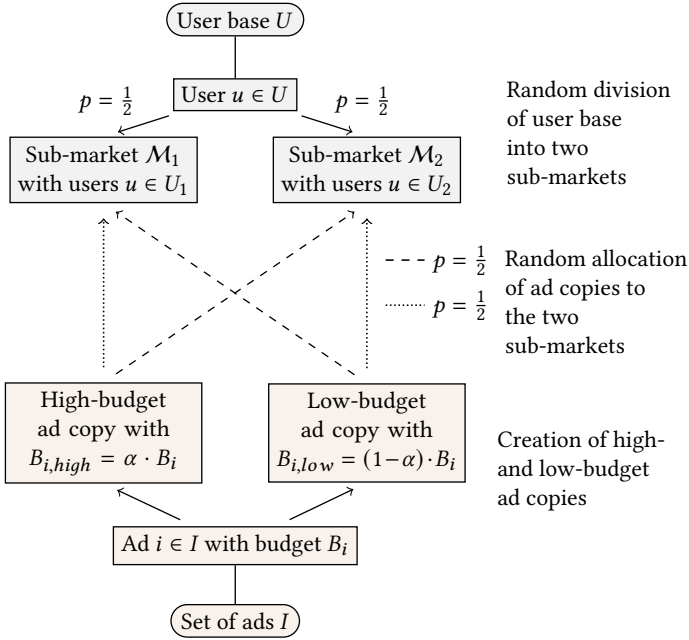

Asymmetric budget split is implemented in three stages: (1) two independent sub-markets are created by randomly partitioning users; (2) each ad is split in two ad copies, with total budget split asymmetrically between the two; (3) each ad copy is randomly assigned to one of the two sub-markets. The full process is described below and visualized in Figure \ref{fig:asym_bs_overview}.

\subsubsection{Randomize users into two distinct sub-markets}\label{sssec:step_1}
Asymmetric budget split follows the regular budget split procedure \cite{liu2020trustworthy,basse2016randomization} by splitting the user base in equally-sized and disjoint sub-markets. Formally, each user $u$ in the full set of users of the ad platform $U$ is with equal probability randomly allocated to one of two disjoint sets of users $U_1$ and $U_2$.
The two sets $U_1$ and $U_2$ form the user base for the two independent ad sub-markets $\mathcal{M}_1$ and $\mathcal{M}_2$ respectively.

\subsubsection{Create two copies of each ad with asymmetric budgets}
Next, the procedure creates two copies of each ad $i$ which inherit all ad features $X_i$ of $i$ except for budget. Instead of dividing budget equally among the two ad copies (or, more generally, in proportion to sub-market size), the method splits an ad's total budget $B_i$ asymmetrically to obtain two distinct copies of type `high' and `low' budget $j\in\{high,\ low\}$. The budget of an ad copy ($i,j$) is given by
\begin{equation*}
    B_{i,j} = 
    \begin{cases}
    \alpha\cdot B_i & \text{if\ }j=high \\
    (1-\alpha)\cdot B_i & \text{if\ }j=low,
    \end{cases}
\end{equation*}
where $\alpha > 0.5$ governs the degree of asymmetry. More generally, different ads can be split with different degrees of asymmetry.

\subsubsection{Randomization of ad budget across the two sub-markets}
High- and low-budget ad copies are then randomly assigned to sub-markets $\mathcal{M}_1$ and $\mathcal{M}_2$ with equal probability $p = 0.5$ (represented by dashed and dotted lines in the bottom half of Figure \ref{fig:asym_bs_overview}). For each ad, this creates two possible allocations of the high- and low-budget copies as shown in Table \ref{tab:allocations}: Allocation I: ($high \rightarrow \mathcal{M}_1;\ low \rightarrow \mathcal{M}_2)$ with $p = 0.5$; and Allocation II: ($high \rightarrow \mathcal{M}_2;\ low \rightarrow \mathcal{M}_1)$ with $p = 0.5$.

\begin{table}[h]
\begin{tabular}{ccc}
  \begin{tabular}{|c|c|c|}
     \multicolumn{1}{c}{} &
     \multicolumn{1}{c}{Sub-market $\mathcal{M}_1$} & \multicolumn{1}{c}{Sub-market $\mathcal{M}_2$}   \\ \cline{2-3}
    \multicolumn{1}{c|}{Low budget}& Allocation I   & Allocation II  \\ \cline{2-3}
    \multicolumn{1}{c|}{High budget} & Allocation II  & Allocation I  \\ \cline{2-3}
  \end{tabular}

\end{tabular}
\caption{Allocation of ad copies across sub-markets}
\label{tab:allocations}
\end{table}

The platform's value delivery mechanisms (such as bidding, pacing, and ranking algorithms) are operated independently across sub-markets $\mathcal{M}_1$ and $\mathcal{M}_2$, in effect creating two entirely separate ad sub-markets. Since allocation of users to $U_1$ and $U_2$ is random, both sub-markets $\mathcal{M}_1$ and $\mathcal{M}_2$ are identical in expectation in terms of their observable and unobservable user characteristics. Random allocation of the high- and low-budget ad copies ensures that $\mathcal{M}_1$ and $\mathcal{M}_2$ are identical in expectation in terms of the budget distribution. Sub-markets are hence comparable.

The procedure to implement the asymmetric budget split methodology is summarized below.
\begin{tcolorbox}[boxrule=0.75pt]
\textbf{Asymmetric Budget Split methodology}
\bigskip

\underline{Step 1:} Randomly divide user base $U$ into two disjoint and equally sized sets $U_1$ and $U_2$ to create ad sub-markets $\mathcal{M}_1$ and $\mathcal{M}_2$. 
\medskip

\underline{Step 2:} For each ad $i$ with budget $B_i$, split its budget asymmetrically by creating a high- and low-budget ad copy ($i,j$) with $j \in \{high, low\}$ and budgets $B_{i,high}=\alpha \cdot B_i$ and $B_{i,low}=(1-\alpha) \cdot B_i$ respectively, where $\alpha > 0.5$.
\medskip

\underline{Step 3:} For each ad $i$, randomly allocate the high- and low-budget copy across $\mathcal{M}_1$ and $\mathcal{M}_2$ such that ($high \rightarrow \mathcal{M}_1;\ low \rightarrow \mathcal{M}_2) \text{ with } p =0.5$ and ($high \rightarrow \mathcal{M}_2;\ low \rightarrow \mathcal{M}_1) \text{ with } p =0.5$.

\end{tcolorbox}

As comparison, the budget-split design of \cite{liu2020trustworthy} is a special case of this procedure with $\alpha = 0.5$ (symmetric split). Step 1 is identical. Step 2 is identical but vacuous, as $\alpha = 0.5$. Step 3 is redundant since the high- and low-budget copies are identical when $\alpha = 0.5$.

Note that to simplify the exposition, we only consider the special case of asymmetric budget split with two equally-sized sub-markets. The methodology can be extended to $n$ equally-sized sub-markets, with an appropriate $n$-fold asymmetric partition of ad budget. It can also be extended to sub-markets of different sizes, where (asymmetric) budgets are scaled according to sub-market size.

\subsection{Incrementality estimation}\label{sec:estimation}
The random variation in budget $B_{i,j}$ across high- and low-budget ad copies induced by asymmetric budget split can be used to obtain a valid estimate of incrementality. The estimation process consists of (1) constructing a random budget feature and (2) inference.

\begin{enumerate}
    \item First, define the random budget component for ad copy ($i,j$) as $RandB_{i,j} = B_{i,j}-0.5 \cdot B_i$. This is equal to \begin{equation*}
        RandB_{i,j} = 
        \begin{cases}
            (\alpha-0.5)\cdot B_i & \text{if\ }j=high \\
            (0.5-\alpha)\cdot B_i & \text{if\ }j=low.
        \end{cases}
    \end{equation*}
    $RandB_{i,j}$ is the deviation from equal split. It isolates the random variation created by asymmetric budget split from other naturally-occurring variation in budget across ads. Note that the distribution of $RandB_{i,j}$ is symmetric and centered around 0 (i.e. $\mathrm{E}[RandB_{i,j}] = 0$) for any $\alpha$.
    
    \item Next, train a regression model that fits 
    performance $Y_{i,j}$ of ad copy ($i,j$) as a function of the random budget component $RandB_{i,j}$ and control features:
    \begin{equation}\label{eq:incrementality_reg}
        Y_{i,j} = cons + \rho \cdot RandB_{i,j} + \xi \times \Lambda_i + \varepsilon_{i,j},
    \end{equation}
    where $\Lambda_{i}$ is an (optional) vector of control features (a subset of ad features $X_i$, such as ad format and targeting criteria). To obtain a population-representative estimate of average incrementality across the ad population, observations are weighted using weights $w = 1/(RandB_{i,j})^2$. To increase statistical power, the vector $\Lambda_i$ may be replaced by a set of fixed effects for $i$. As error terms are correlated between the two copies of an ad $i$, standard errors are clustered at the ad level for valid inference. We discuss interpretation of \eqref{eq:incrementality_reg} and expand on the observations above in the remainder of this section.
\end{enumerate}

In regression model \eqref{eq:incrementality_reg}, the coefficient
$\hat{\rho}$ is  
an estimate of the average treatment effect of a one-unit increase in budget on expected ad performance $Y$ across all ads. If the analyst is interested in heterogeneous estimates for different ad segments (such as by country or by ad type), model (\ref{eq:incrementality_reg}) can be estimated separately for each ad segment.

Note that variation in $RandB_{i,j}$ is larger for ad copies whose parent ad $i$ has a larger $B_i$. An unweighted regression estimate of $\hat{\rho}$ would therefore load more strongly on ads with larger budgets. 
Using weights $w$ as defined above ensures that each ad copy contributes equally. This ensures that $\hat{\rho}$ is a population-representative estimate of average incrementality across the entire marketplace.

Turning to the statistical precision of the incrementality estimate $\hat{\rho}$, two aspects of regression model \eqref{eq:incrementality_reg} are noteworthy. First, note that each ad $i$ is represented twice in the dataset in form of a high- and a low-budget ad copy. Hence, incrementality can be identified from variation in budget \textit{within ads} only. In particular, if ad fixed effects are included in model \eqref{eq:incrementality_reg}, all ad-specific, idiosyncratic variation in performance is absorbed across ads. This substantially increases the statistical precision in the incrementality estimate $\hat{\rho}$, particularly when there are other (potentially unobservable) ad-specific factors besides budget that have a large contribution to the overall variance in outcomes. 

Second, the statistical precision of the incrementality estimate $\hat{\rho}$ critically depends on the variance of the random budget component $RandB_{i,j}$ which is controlled by asymmetric split factor $\alpha$. The larger the degree of asymmetry, the higher the statistical precision of the incrementality estimate. The ad platform can perform power calculations to adjust the level of $\alpha$ in order to reach the desired level of statistical precision.

\section{Data and system architecture}\label{sec:architecture}
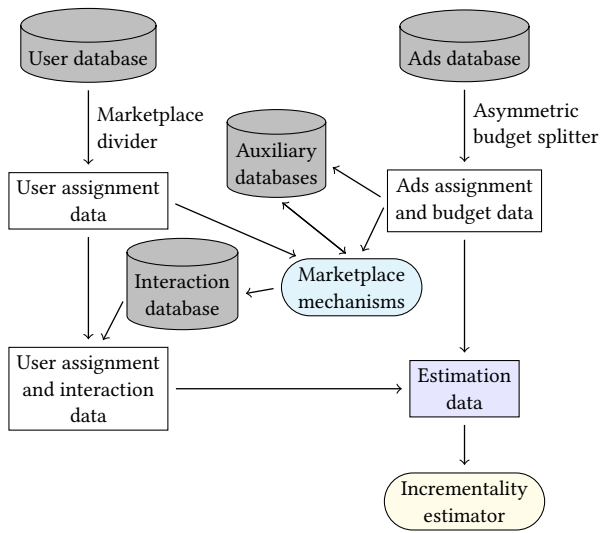
\begin{figure}[h]
\begin{tikzpicture}[font=\small, database/.style={
      cylinder,
      cylinder uses custom fill,
      cylinder body fill=gray!50,
      cylinder end fill=gray!50,
      shape border rotate=90,
      aspect=0.25,
      draw
    }]

\node[database] (userdata) at (-2,0.4) {User database};
\node[database] (adsdata) at (3,0.4) {Ads database};

\node[draw, align=center] (userassignment) at (-2,-1.5) {User assignment\\ data};
\node[draw, align=center] (adsassignment) at (3,-1.5) {Ads assignment\\ and budget data};

\node[database, align=center] (interactiondata) at (-0.75,-2.75) {Interaction\\ database};

\node[draw, align=center] (userinteractiondata) at (-2,-4) {User assignment\\ and interaction\\ data};

\node[draw, align=center,fill=blue!10] (estimationdata) at (3,-4) {Estimation\\ data};

\node[draw, align=center,fill=yellow!10, rounded rectangle] (estimator) at (3,-5.5) {Incrementality\\ estimator};

\draw[align=left, shorten >=0.1cm,shorten <=0.1cm,->, line width=0.5pt] (userdata.south) -- (userassignment.north) node[midway, right] {Marketplace\\ divider};

\draw[align=left, shorten >=0.1cm,shorten <=0.1cm,->, line width=0.5pt] (adsdata.south) -- (adsassignment.north) node[midway, right] {Asymmetric\\ budget splitter};

\draw[shorten >=0.1cm,shorten <=0.1cm,->, line width=0.5pt] (interactiondata.west) -- (userinteractiondata.75) ;

\draw[shorten >=0.1cm,shorten <=0.1cm,->, line width=0.5pt] (userassignment.south) -- (userinteractiondata.north);

\draw[shorten >=0.1cm,shorten <=0.1cm,->, line width=0.5pt] (adsassignment.south) -- (estimationdata.north);

\draw[shorten >=0.1cm,shorten <=0.1cm,->, line width=0.5pt] (userinteractiondata.east) -- (estimationdata.west);

\draw[shorten >=0.1cm,shorten <=0.1cm,->, line width=0.5pt] (estimationdata.south) -- (estimator.north);

\node[draw, rounded rectangle, align=center, fill=cyan!10] (marketplacevalue) at (1.5,-2.65) {Marketplace\\ mechanisms};

\draw [shorten >=0.1cm,shorten <=0.1cm,->, line width=0.5pt] 
    (userassignment.east) -- (marketplacevalue.150) ;

\draw [shorten >=0.1cm,shorten <=0.1cm,->, line width=0.5pt] 
    (adsassignment.west) -- (marketplacevalue.75) ;

\node[database, align=center] (auxiliarydata) at (0.5,-1) {Auxiliary\\ databases};

\draw [shorten >=0.1cm,shorten <=0.1cm,->, line width=0.5pt] 
    (adsassignment.west) -- (auxiliarydata.east) ;

\draw [shorten >=0.1cm,shorten <=0.1cm,->, line width=0.5pt] 
    (marketplacevalue.west) -- (interactiondata.east) ;

\draw[shorten >=0.1cm,shorten <=0.1cm,->, line width=0.5pt] (auxiliarydata.south) -- (marketplacevalue.north);
\draw[shorten >=0.1cm,shorten <=0.1cm,<-, line width=0.5pt] (auxiliarydata.south) -- (marketplacevalue.north);

\end{tikzpicture}

\caption{Data and system architecture}

\label{fig:data_architecture}

\end{figure}

Figure \ref{fig:data_architecture} shows the data and system architecture that supports asymmetric budget split. The primary system input data are stored in three databases:\begin{enumerate}
\item[1.] The \textit{user database} contains all platform users $u\in U$ with user identifiers. This only includes user ID.
\item[2.] The \textit{ads database} contains all ads $i\in I$, including their identifiers, budget information $B_i$, and ad features $X_i$.
\item[3.] The \textit{interaction database} contains all user-ad interactions $l \in L$, including the associated user id $u(l)$ and ad identifier $i(l)$ and type of interaction (click, conversion, etc).
\end{enumerate}

Data from the user database is passed to the \textit{marketplace divider}, which randomly divides users into two sub-markets $\mathcal{M}_1$ and $\mathcal{M}_2$; the resulting assignment is stored as intermediate \textit{user assignment data}. 

Data from the ads database is passed to the \textit{asymmetric budget splitter}, which creates high- and low-budget ad copies for each ad, and randomly assigns each copy to one sub-market $\mathcal{M}_1$ or $\mathcal{M}_2$. The assignment and asymmetric budget information is stored as intermediate \textit{ads assignment and budget data}, which contains two rows per ad (one for the high-budget copy and one for the low-budget copy). Both copies inherit ad features $X_i$.

Marketplace mechanisms use the user assignment data as well as the ads assignment and budget data as inputs. Based on this data and additional data from auxiliary databases,\footnote{The auxiliary databases store any additional data needed by system sub-components that perform marketplace mechanisms. For instance, this may include additional features on users and their activity which is used to predict ads relevance. It also includes real-time market information such as remaining budget for every ad copy. Real-time market information is updated based on the results of marketplace mechanisms.}
marketplace mechanisms determine the ads shown to the user given a certain request. Depending on the application, this may include determining which ad or set of ads to display, and in which order; their position; and their size. The specific nature of marketplace mechanisms depends on the application. Modern ad platforms typically employ auction-based systems which rely on components such as user interaction prediction models, auto-bidding, and budget pacing. Independent of the specific system, upon an impression of an ad, the user interaction and its type (such as a click or a conversion) are recorded and stored in the user interaction database, which contains one row per user interaction.

To generate the estimation data, the following data processing is required: first, user interaction data is joined with the user assignment data based on user identifier $u$. The individual user interactions are then aggregated to create a performance measure $Y_{i,j}$ for each ad copy in its respective sub-market. Next, the resulting intermediate dataset is joined with the ads assignment and budget data using ad identifier $i$ and an indicator for the sub-market. The resulting \textit{estimation data} contains one row per ad copy $(i,j)$ and stores information on ad performance in the respective sub-market $Y_{i,j}$, the associated budget level $B_{i,j}$, and inherited ad features $X_i$ as well as $B_i$.

Finally, the estimation data is passed to the the \textit{incrementality estimator} which performs incrementality estimation, as described in section \ref{sec:estimation}.

\section{Application: LinkedIn Jobs Marketplace}\label{sec:application}
\subsection{Marketplace description}
Asymmetric budget split has been succesfully deployed to LinkedIn's Jobs Marketplace to reliably measure incrementality. LinkedIn's Jobs Marketplace connects job \textit{seekers} (users, in the language used in the rest of the paper) with job \textit{posters} (advertisers).

Job seekers may be registered LinkedIn members or ``guest'' seekers in a logged-out state, and range from active job hunters to passive candidates. The platform offers multiple job seeking tools, from Job Search (\href{https://www.linkedin.com/jobs/search/}{linkedin.com/jobs/search}) which allows job seekers to specify keywords and filter results (by seniority level, location, company, etc), to job recommendations (\href{https://www.linkedin.com/jobs/}{linkedin.com/jobs}) that the platform suggests based on seekers' profile and stated preferences.

Job posters can list their job openings in a variety of ways, from free listings to subscription-like solutions to pay-for-performance offerings. The last category is of special interest to this paper. Job posters can \textit{promote} their free job listings (\href{https://www.linkedin.com/talent/post-a-job}{linkedin.com/talent/post-a-job}) in order to receive additional visibility. This may come in the form of, among others, access to exclusive channels (email, mobile push notifications), prominent positioning and visual decorations (``Promoted'') in non-exclusive channels, and ranking advantage within promoted placements. For each promoted job listing $i$, its job poster specifies its budget $B_i$ to be used for promotion,\footnote{For some job products, the platform assigns budgets on behalf of job posters.} either as a daily budget or throughout the lifetime of the job.

Thanks to the asymmetric budget split methodology introduced in this paper, LinkedIn has obtained valid estimates of incremental returns of job posting budget $B_i$ in terms of outcomes of interest to job posters. Such outcomes include how many times a job listing is viewed by seekers and how many applicants it receives (the natural notion of conversion in this context). The platform-controlled asymmetric split factor $\alpha$ was set to $\alpha=0.6$ using a power calculation that considers the overall variation in job outcomes across the marketplace. 

Measuring incrementality is of primary interest to job posters as it informs and guides their choice of budget. Estimates of incrementality are used as inputs to performance forecasting tools that are provided to job posters, as well to validate the outputs of these tools. Incrementality is also of direct interest to LinkedIn: it is used to understand the dynamics and health of the marketplace, as well as to prioritize among different initiatives to improve its marketplace mechanisms. Because of the value of up-to-date incrementality estimates, asymmetric budget split is deployed continuously, not as a one-off measurement exercise.

\subsection{Incrementality estimation}
We present incrementality estimates for job posters on LinkedIn's Jobs Marketplace obtained using asymmetric budget split. As job posters promote their job listings in order to attract a higher number of applicants, we focus the presentation on the number of applicants as the key outcome for advertisers. Specifically, we present estimates of how many additional applicants the average job poster can expect in return for a one unit higher budget on their job posting. As a common denomination we express budgets in terms of US dollars.

\begin{figure}[h]
  \centering
  \includegraphics[width=\linewidth]{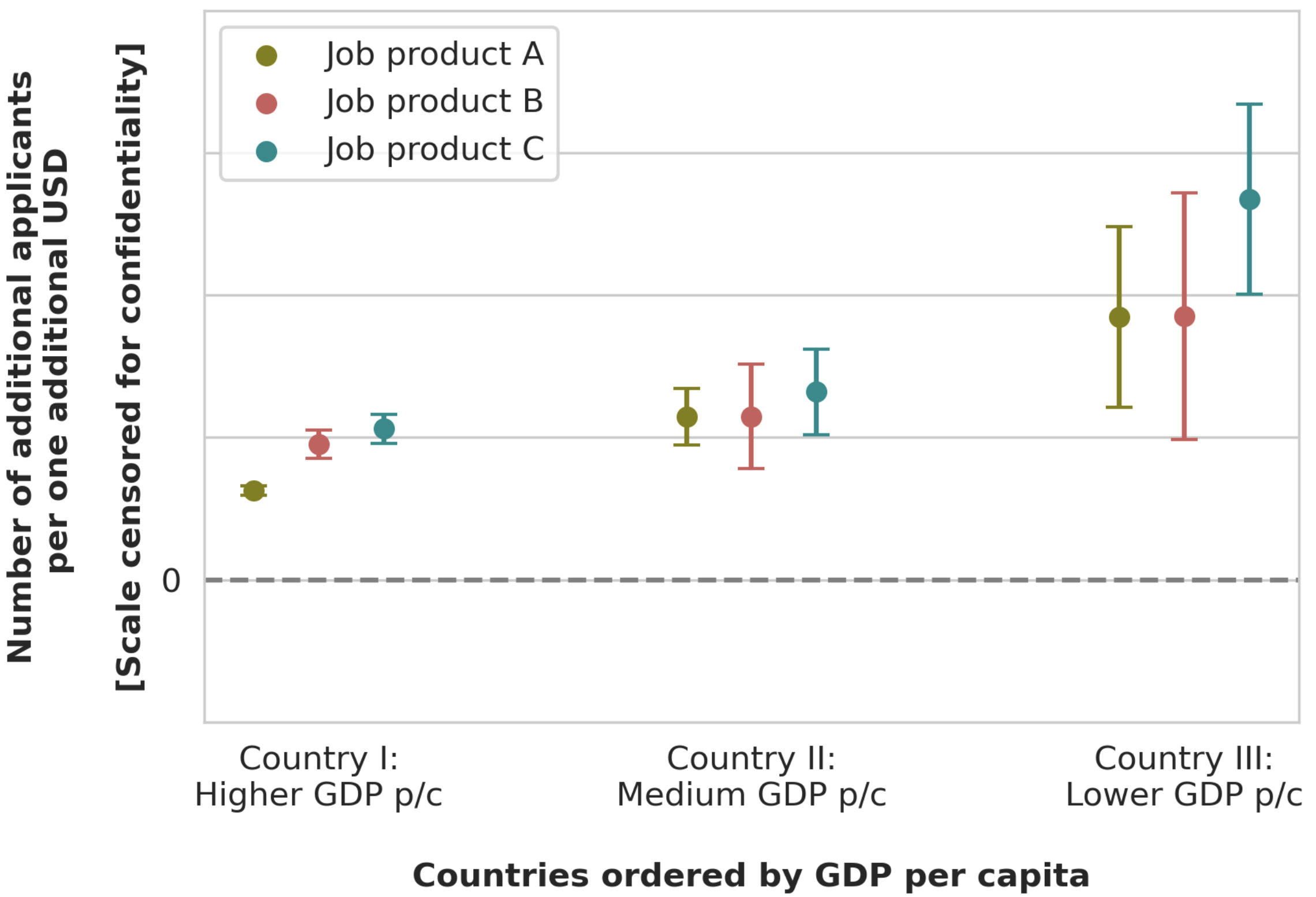}
  \caption{Incremental return estimates by country. The figure shows the estimated number of additional applicants per one US dollar of additional budget for three job products and three countries (ordered by GDP per capita). Error bars indicate 95\%-confidence intervals. 
  }
    \label{fig:incrementality_by_country}
\end{figure}

Figure \ref{fig:incrementality_by_country} shows estimates of incremental returns for three different job products and three different countries (ordered by their GDP per capita). Error bars indicate 95\%-confidence intervals. For reasons of confidentiality, the y-axis is censored. 

Incremental returns are comparable for the three different job products. However, the estimates indicate substantial heterogeneity across countries. In particular, the expected number of additional applicants delivered in return for one additional unit of budget is lower in countries with a larger GDP per capita. This pattern follows natural market dynamics: in higher-income countries job posters' willingness to pay for applicants is higher, leading to a higher market price, and as a consequence lower incremental returns per one additional unit of budget.

\begin{figure}[h]
  \centering
  \includegraphics[width=\linewidth]{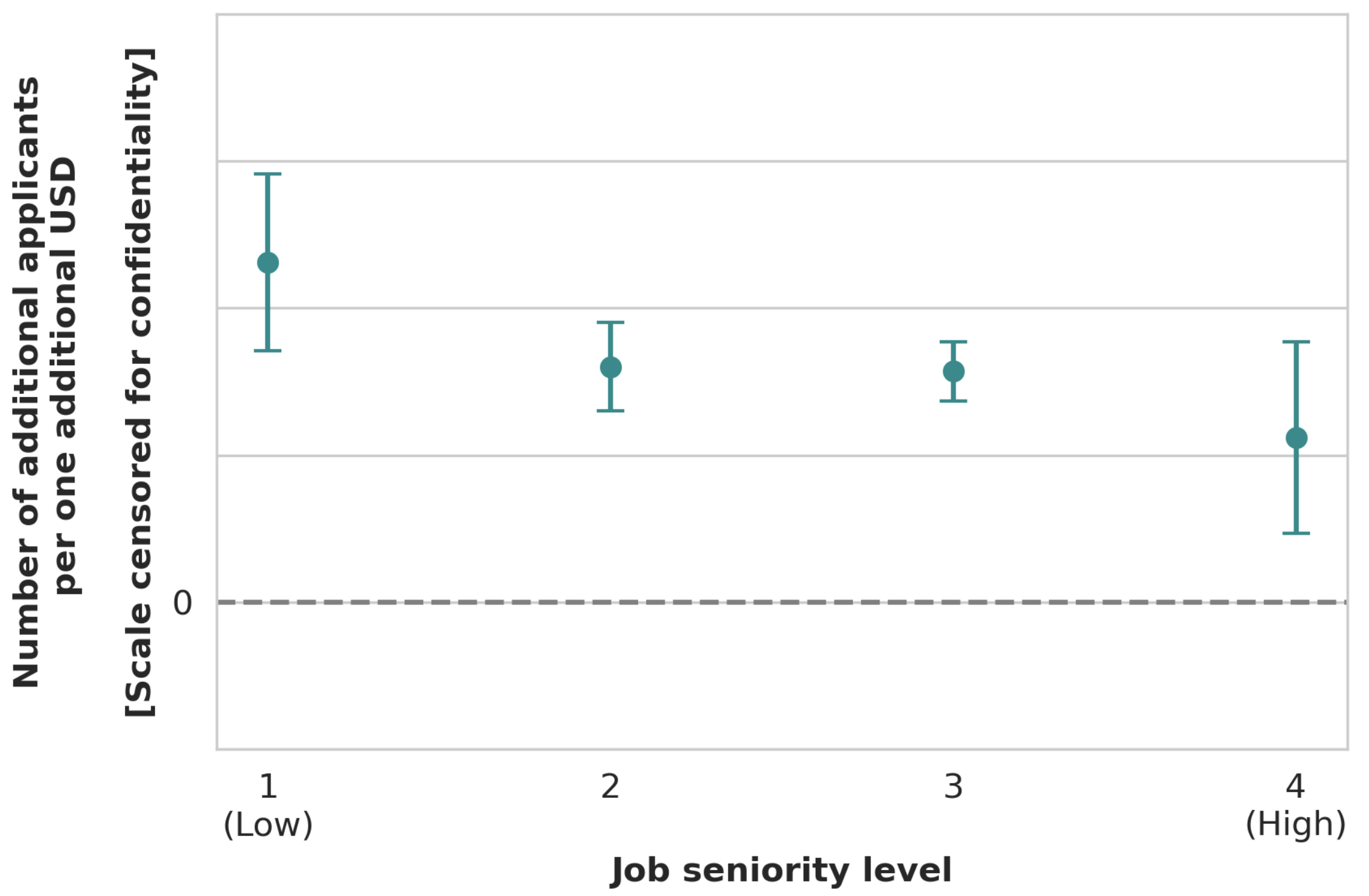}
  \caption{Incremental return estimates by job seniority level. The figure shows the estimated number of additional applicants per one US dollar additional budget for jobs with different seniority level (for one selected job product and country). Error bars indicate 95\%-confidence intervals. 
  }
    \label{fig:incrementality_by_seniority}
\end{figure}

As a further dimension of heterogeneity, we show in Figure \ref{fig:incrementality_by_seniority} the heterogeneous incremental returns for job postings with different seniority levels. We observe that the expected number of additional applicants delivered in return for one additional unit of budget is lower for job postings that require a higher candidate seniority level. Again, this pattern follows natural market dynamics: job posters have a higher willingness to pay for more senior candidates, of which supply is scarcer. As a consequence, for more senior candidates the market price is higher, and hence incremental returns per one additional unit of budget are lower.

\section{Operational considerations and further use cases}\label{sec:considerations}

In this section we outline practical considerations relevant for ad platforms that are considering implementing asymmetric budget split, as well as additional use cases beyond measurement of advertising returns. 

\subsection{Operational considerations}\label{sec:operational}
Asymmetric budget split is particularly attractive to implement for ad platforms that already have the infrastructure for regular (symmetric) budget split \cite{liu2020trustworthy}. Although the purpose of the two methods is entirely distinct, the only significant additional engineering effort required is to enable \textit{asymmetric} splits. 

Switching a marketplace from symmetric budget split $(\alpha=0.5)$ to asymmetric split $(\alpha>0.5)$ may have side effects on its operation. We discuss these effects in terms of their consequences for the performance of individual ads as well as the marketplace as whole. In general, potential side effects are smaller in marketplaces (1) with a large number of ads, (2) where any individual ad budget is small relative to the total budget, and (3) where individual budgets are small relative to the targeted population. Whether the platform values valid incrementality estimates more or less than the cost of the side effects is application-specific. This trade-off must also be evaluated when deciding the degree of asymmetry $\alpha$: the farther $\alpha$ is from $0.5$, the greater the statistical power, but the greater the potential side effects. We recommend conducting a power analysis to determine statistical precision of estimated incrementality as a function of $\alpha$; and then choosing the $\alpha$ closest to $0.5$ that meets the desired level.

\subsubsection{Performance of individual ads}
When an ad's budget is split asymmetrically its performance may be lower than what it would have been under symmetric budget split. In the basic asymmetric budget split design, the high-budget and low-budget ad copies both target equally-sized user populations. When there are diminishing returns (for example if performance is a concave increasing function of budget), expected performance is maximized when budgets are split in proportion to sub-market size. Performance here is meant at the level of the entire ad; differential performance of ad copies across sub-markets is intended, and is what allows incrementality estimation. While in many applications diminishing returns are expected, what matters in practice is the degree of diminishing returns in the (local) region $(1-\alpha)B_i$ to $\alpha B_i$. When $\alpha$ is close to $0.5$ and ad budgets are small relative to the targeted population, performance is roughly linear in budget (locally) and hence this side effect is negligible.

This effect can be sized by running an ad-level A/B test: a random subset of ads are split with $\alpha_1 > 0.5$; other ads are split with $\alpha_0 = 0.5$ (i.e. symmetric split); then average performance can be compared at different levels of $\alpha$ to quantify the performance impact of asymmetric budget split. For our application to LinkedIn's Job Marketplace we conducted such ad-level A/B tests, comparing ads with asymmetrically and symmetrically split budgets, and found no economically or statistically significant impact on performance.  

\subsubsection{Marketplace performance}
A second type of performance side effect may occur at the sub-market level. By switching from symmetric to asymmetric budget split, the new distribution of budgets (loosely) second-order stochastically dominates the original distribution. (This statement is loose because the distribution is itself random due to the random allocation of high-budget and low-budget copies across the two sub-markets.) Average budget is unchanged, but the variance increases (the more so the farther the degree of asymmetry $\alpha$ is from $0.5$). Depending on the specific marketplace mechanisms in place, this shift may affect overall performance of the marketplace. This side effect can be tested for through switchback (or alternate day) test designs \cite{bojinov2020design}.

\subsubsection{Validity of experimental platform}
The budget-split design \cite{liu2020trustworthy} is a keystone of marketplace experimentation. It is imperative that the adaptations required by asymmetric budget split do not compromise the validity of marketplace experiments run with a budget-split design. In such experiments, different marketplace mechanisms are implemented in each sub-market, and their impact is measured by comparing sub-markets.  

With (symmetric) budget split, sub-markets are exactly identical on the ad side: each budget is split equally, in proportion to the sub-market size. Any imbalance comes from randomization on the user side, as for all A/B tests. With asymmetric budget split, sub-markets also differ on the ad side in that, by design, individual ad budgets are split asymmetrically. However, since the high- and low-budget ad copies are randomly assigned to sub-markets, the budget distribution is the same in expectation across the sub-markets. This ensures that the integrity of the budget-split design for marketplace experimentation remains unaffected.

To verify that asymmetric budget split does not invalidate the balance across sub-markets in practice, an analyst may check that moments of the budget distribution (such as mean and variance) are comparable across sub-markets, both in aggregate and by sub-populations. This is particularly advisable if the number of ads in the entire market is small. By the law of large numbers, balance will be guaranteed when there are many ads and each individual budget is small relative to the total size. It is also advisable for an ad platform to conduct such balance tests when deploying asymmetric budget split for the first time, in order to verify that the asymmetric split and randomization across sub-markets are correctly implemented in the system.
 
\subsection{Further use cases}
\subsubsection{Measuring incrementality lift in A/B tests}\label{sssec:incrementality_impact}
Besides measuring incrementality in their current value delivery system, ad platforms may want to assess the impact on incrementality when testing a change to their system, such as changes in auto-bidding or ranking mechanisms. This is of particular importance when the change being tested aims to improve returns for advertisers.

For such assessments, asymmetric budget split can be combined with regular budget split for marketplace experimentation through A/B tests \cite{tang2010overlapping,kohavi2012trustworthy,deng2016data}. Consider two value delivery mechanisms, control $C$ (the current mechanism) and treatment $T$ (a candidate mechanism). As in regular budget split \cite{liu2020trustworthy}, mechanism $C$ is assigned to sub-market $\mathcal{M}_1$ while mechanism $T$ is assigned to sub-market $\mathcal{M}_2$ (without loss of generality). Mechanisms $C$ and $T$ operate independently in the two sub-markets, shutting down any interference due to budget constraints. 

Using this strategy, the incrementality lift of value delivery mechanism $T$ relative to $C$ can be measured using a regression model of the form:
\begin{equation}\label{eq:incrementality_ab_test}
\begin{split}
Y_{i,j} &= cons + \mu \cdot m_{i,j} + \rho \cdot RandB_{i,j} \\ &\hphantom{= x} + \Delta\rho \cdot RandB_{i,j} \cdot m_{i,j} +   \xi \times \Lambda_i + \varepsilon_{i,j},
\end{split}
\end{equation}
where $m_{i,j}$ is a binary indicator for whether ad copy ($i,j$) is assigned to sub-market $\mathcal{M}_2$ (i.e. the treatment market) and all other notation follows the notation of model (\ref{eq:incrementality_reg}). Standard errors as well as weights are calculated as for model (\ref{eq:incrementality_reg}).

In regression model (\ref{eq:incrementality_ab_test}), the estimated coefficient $\hat{\rho}$ represents advertising returns (i.e. incrementality) under the control value delivery mechanism $C$. The estimated coefficient $\hat{\mu}$ captures the average budget-independent ad performance impact of the treatment value delivery mechanism $T$ relative to control $C$. Finally, the estimated coefficient $\widehat{\Delta\rho}$ is the parameter of interest: it captures the causal impact of value delivery mechanism $T$ on incrementality relative to control mechanism $C$.

\subsubsection{Performance forecasting}
Asymmetric budget split can also be leveraged for performance forecasting \cite{nath2013ad,parmar2017forecasting}.

Advertisers increasingly demand forecasts regarding the performance of their ads, for instance in terms of the expected number of clicks or conversions. For an informed investment decision, it is critical for advertisers to know expected performance at different ad budget levels. To address this need, ad platforms commonly provide this information to advertisers through performance forecasts that change in response to the budget selection of the advertiser.

Such budget-aware performance forecasting models trained on historical data, however, face the same causal inference challenges as causal measurements of incrementality: budget is typically non-random and correlated with other unobservable ad-specific features. This can lead to omitted variable and selection bias. Hence, when trained on historical data, the model-implied relationship between budget and  performance can be severely biased. 

Asymmetric budget split solves this challenge by providing training data in which budget variation is orthogonal to other ad-specific features. Akin to equation \eqref{eq:incrementality_reg}, forecasting models can be trained on data from asymmetric budget split in a way that the relationship between budget and forecasts is only identified from randomly occurring variation in budget.
Hence, when the model is scored at different budget levels, the implied change in ad performance matches the causal effect of changing budgets. This allows the platform to provide accurate budget-aware performance forecasts to advertisers. 

\section{Conclusion}\label{sec:conclusion}
We introduce asymmetric budget split, a novel method for ad platforms to measure advertising returns in an unobtrusive manner. Asymmetric budget split induces small random perturbations in ad budget allocation across separate and comparable sub-markets of the platform. This enables the platform to observe performance of the same ad at different budget levels holding other factors constant. Assessing the change in ad performance in response to the differences in budget allows the platform to reliably and causally measure advertising returns. 

Asymmetric budget split has three properties that make it particularly appealing for ad platforms to implement. First, estimates of advertising returns from asymmetric budget split are causally valid. Relative to observational methods that utilize correlational relationships in historical data, asymmetric budget split provides exogenous (i.e. random) variation in budget. Second, asymmetric budget split is unobtrusive. The method does not alter any ads' total budget on the platform. Nor does it change the rate at which ads are impressed nor are users hold out from seeing an ad. Any ad can potentially be displayed to any user. As such, effects on user experience are minimal; in addition, as long as asymmetries in budget allocation are small, ad performance on the platform is not hurt. Third and related, asymmetric budget split is cost-effective. As no holdout groups are required and given budget asymmetries are small, operational costs for the ad platform are low. This allows ad platforms to employ asymmetric budget split on an ongoing basis for efficient performance management and monitoring of marketplace health.

Besides accurately measuring advertising returns, asymmetric budget split has a number of additional use cases. In combination with regular budget split, the method can be used to measure incrementality lifts in marketplace experiments. Moreover, it provides a source of random variation in budget which can be leveraged as training data for causally-valid budget-aware performance forecasting models.

\begin{acks}
The authors would like to thank Tilbe \c{C}a\u{g}layan and Shan Zhou for their partnership in deploying asymmetric budget split at LinkedIn; Min Liu, Kay Shen, and Rose Tan for reviewing the paper; and Noureddine El Karoui, Linda Fayad, Min Liu, Di Luo, Chinmay Kini, Cindy Liang, Alex Patry, Wen Pu, Suju Rajan, Zheng Shao, Jerry Shen, Kay Shen, Dawei Wang, and Junyu Yang for valuable feedback at all stages of the project.
\end{acks}

\bibliographystyle{ACM-Reference-Format}
\bibliography{sample-base}


\begin{thebibliography}{29}


\ifx \showCODEN    \undefined \def \showCODEN     #1{\unskip}     \fi
\ifx \showDOI      \undefined \def \showDOI       #1{#1}\fi
\ifx \showISBNx    \undefined \def \showISBNx     #1{\unskip}     \fi
\ifx \showISBNxiii \undefined \def \showISBNxiii  #1{\unskip}     \fi
\ifx \showISSN     \undefined \def \showISSN      #1{\unskip}     \fi
\ifx \showLCCN     \undefined \def \showLCCN      #1{\unskip}     \fi
\ifx \shownote     \undefined \def \shownote      #1{#1}          \fi
\ifx \showarticletitle \undefined \def \showarticletitle #1{#1}   \fi
\ifx \showURL      \undefined \def \showURL       {\relax}        \fi
\providecommand\bibfield[2]{#2}
\providecommand\bibinfo[2]{#2}
\providecommand\natexlab[1]{#1}
\providecommand\showeprint[2][]{arXiv:#2}

\bibitem[Agarwal et~al\mbox{.}(2014)]%
        {agarwal2014budget}
\bibfield{author}{\bibinfo{person}{Deepak Agarwal}, \bibinfo{person}{Souvik
  Ghosh}, \bibinfo{person}{Kai Wei}, {and} \bibinfo{person}{Siyu You}.}
  \bibinfo{year}{2014}\natexlab{}.
\newblock \showarticletitle{Budget pacing for targeted online advertisements at
  linkedin}. In \bibinfo{booktitle}{\emph{Proceedings of the 20th ACM SIGKDD
  international conference on Knowledge discovery and data mining}}.
  \bibinfo{pages}{1613--1619}.
\newblock


\bibitem[Basse et~al\mbox{.}(2016)]%
        {basse2016randomization}
\bibfield{author}{\bibinfo{person}{Guillaume~W Basse},
  \bibinfo{person}{Hossein~Azari Soufiani}, {and} \bibinfo{person}{Diane
  Lambert}.} \bibinfo{year}{2016}\natexlab{}.
\newblock \showarticletitle{Randomization and the pernicious effects of limited
  budgets on auction experiments}. In \bibinfo{booktitle}{\emph{Artificial
  Intelligence and Statistics}}. PMLR, \bibinfo{pages}{1412--1420}.
\newblock


\bibitem[Bojinov et~al\mbox{.}(2020)]%
        {bojinov2020design}
\bibfield{author}{\bibinfo{person}{Iavor Bojinov}, \bibinfo{person}{David
  Simchi-Levi}, {and} \bibinfo{person}{Jinglong Zhao}.}
  \bibinfo{year}{2020}\natexlab{}.
\newblock \showarticletitle{Design and Analysis of Switchback Experiments}.
\newblock \bibinfo{journal}{\emph{arXiv preprint arXiv:2009.00148}}
  (\bibinfo{year}{2020}).
\newblock


\bibitem[Deng and Shi(2016)]%
        {deng2016data}
\bibfield{author}{\bibinfo{person}{Alex Deng} {and} \bibinfo{person}{Xiaolin
  Shi}.} \bibinfo{year}{2016}\natexlab{}.
\newblock \showarticletitle{Data-driven metric development for online
  controlled experiments: Seven lessons learned}. In
  \bibinfo{booktitle}{\emph{Proceedings of the 22nd ACM SIGKDD International
  Conference on Knowledge Discovery and Data Mining}}. \bibinfo{pages}{77--86}.
\newblock


\bibitem[Evans(2009)]%
        {evans2009online}
\bibfield{author}{\bibinfo{person}{David~S Evans}.}
  \bibinfo{year}{2009}\natexlab{}.
\newblock \showarticletitle{The online advertising industry: Economics,
  evolution, and privacy}.
\newblock \bibinfo{journal}{\emph{Journal of economic perspectives}}
  \bibinfo{volume}{23}, \bibinfo{number}{3} (\bibinfo{year}{2009}),
  \bibinfo{pages}{37--60}.
\newblock


\bibitem[Farahat and Bailey(2012)]%
        {farahat2012effective}
\bibfield{author}{\bibinfo{person}{Ayman Farahat} {and}
  \bibinfo{person}{Michael~C Bailey}.} \bibinfo{year}{2012}\natexlab{}.
\newblock \showarticletitle{How effective is targeted advertising?}. In
  \bibinfo{booktitle}{\emph{Proceedings of the 21st international conference on
  World Wide Web}}. \bibinfo{pages}{111--120}.
\newblock


\bibitem[Goldfarb and Tucker(2011)]%
        {goldfarb2011online}
\bibfield{author}{\bibinfo{person}{Avi Goldfarb} {and}
  \bibinfo{person}{Catherine Tucker}.} \bibinfo{year}{2011}\natexlab{}.
\newblock \showarticletitle{Online display advertising: Targeting and
  obtrusiveness}.
\newblock \bibinfo{journal}{\emph{Marketing Science}} \bibinfo{volume}{30},
  \bibinfo{number}{3} (\bibinfo{year}{2011}), \bibinfo{pages}{389--404}.
\newblock


\bibitem[Goldfarb and Tucker(2019)]%
        {goldfarb2019digital}
\bibfield{author}{\bibinfo{person}{Avi Goldfarb} {and}
  \bibinfo{person}{Catherine Tucker}.} \bibinfo{year}{2019}\natexlab{}.
\newblock \showarticletitle{Digital economics}.
\newblock \bibinfo{journal}{\emph{Journal of Economic Literature}}
  \bibinfo{volume}{57}, \bibinfo{number}{1} (\bibinfo{year}{2019}),
  \bibinfo{pages}{3--43}.
\newblock


\bibitem[Gordon et~al\mbox{.}(2021)]%
        {gordon2021inefficiencies}
\bibfield{author}{\bibinfo{person}{Brett~R Gordon}, \bibinfo{person}{Kinshuk
  Jerath}, \bibinfo{person}{Zsolt Katona}, \bibinfo{person}{Sridhar Narayanan},
  \bibinfo{person}{Jiwoong Shin}, {and} \bibinfo{person}{Kenneth~C Wilbur}.}
  \bibinfo{year}{2021}\natexlab{}.
\newblock \showarticletitle{Inefficiencies in digital advertising markets}.
\newblock \bibinfo{journal}{\emph{Journal of Marketing}} \bibinfo{volume}{85},
  \bibinfo{number}{1} (\bibinfo{year}{2021}), \bibinfo{pages}{7--25}.
\newblock


\bibitem[Gordon et~al\mbox{.}(2019)]%
        {gordon2019comparison}
\bibfield{author}{\bibinfo{person}{Brett~R Gordon}, \bibinfo{person}{Florian
  Zettelmeyer}, \bibinfo{person}{Neha Bhargava}, {and} \bibinfo{person}{Dan
  Chapsky}.} \bibinfo{year}{2019}\natexlab{}.
\newblock \showarticletitle{A comparison of approaches to advertising
  measurement: Evidence from big field experiments at Facebook}.
\newblock \bibinfo{journal}{\emph{Marketing Science}} \bibinfo{volume}{38},
  \bibinfo{number}{2} (\bibinfo{year}{2019}), \bibinfo{pages}{193--225}.
\newblock


\bibitem[Hoban and Bucklin(2015)]%
        {hoban2015effects}
\bibfield{author}{\bibinfo{person}{Paul~R Hoban} {and}
  \bibinfo{person}{Randolph~E Bucklin}.} \bibinfo{year}{2015}\natexlab{}.
\newblock \showarticletitle{Effects of internet display advertising in the
  purchase funnel: Model-based insights from a randomized field experiment}.
\newblock \bibinfo{journal}{\emph{Journal of Marketing Research}}
  \bibinfo{volume}{52}, \bibinfo{number}{3} (\bibinfo{year}{2015}),
  \bibinfo{pages}{375--393}.
\newblock


\bibitem[Hu et~al\mbox{.}(2016)]%
        {hu2016incentive}
\bibfield{author}{\bibinfo{person}{Yu Hu}, \bibinfo{person}{Jiwoong Shin},
  {and} \bibinfo{person}{Zhulei Tang}.} \bibinfo{year}{2016}\natexlab{}.
\newblock \showarticletitle{Incentive problems in performance-based online
  advertising pricing: Cost per click vs. cost per action}.
\newblock \bibinfo{journal}{\emph{Management Science}} \bibinfo{volume}{62},
  \bibinfo{number}{7} (\bibinfo{year}{2016}), \bibinfo{pages}{2022--2038}.
\newblock


\bibitem[Johnson et~al\mbox{.}(2017a)]%
        {johnson2017ghost}
\bibfield{author}{\bibinfo{person}{Garrett~A Johnson},
  \bibinfo{person}{Randall~A Lewis}, {and} \bibinfo{person}{Elmar~I
  Nubbemeyer}.} \bibinfo{year}{2017}\natexlab{a}.
\newblock \showarticletitle{Ghost ads: Improving the economics of measuring
  online ad effectiveness}.
\newblock \bibinfo{journal}{\emph{Journal of Marketing Research}}
  \bibinfo{volume}{54}, \bibinfo{number}{6} (\bibinfo{year}{2017}),
  \bibinfo{pages}{867--884}.
\newblock


\bibitem[Johnson et~al\mbox{.}(2017b)]%
        {johnson2017less}
\bibfield{author}{\bibinfo{person}{Garrett~A Johnson},
  \bibinfo{person}{Randall~A Lewis}, {and} \bibinfo{person}{David~H Reiley}.}
  \bibinfo{year}{2017}\natexlab{b}.
\newblock \showarticletitle{When less is more: Data and power in advertising
  experiments}.
\newblock \bibinfo{journal}{\emph{Marketing Science}} \bibinfo{volume}{36},
  \bibinfo{number}{1} (\bibinfo{year}{2017}), \bibinfo{pages}{43--53}.
\newblock


\bibitem[Kohavi et~al\mbox{.}(2012)]%
        {kohavi2012trustworthy}
\bibfield{author}{\bibinfo{person}{Ron Kohavi}, \bibinfo{person}{Alex Deng},
  \bibinfo{person}{Brian Frasca}, \bibinfo{person}{Roger Longbotham},
  \bibinfo{person}{Toby Walker}, {and} \bibinfo{person}{Ya Xu}.}
  \bibinfo{year}{2012}\natexlab{}.
\newblock \showarticletitle{Trustworthy online controlled experiments: Five
  puzzling outcomes explained}. In \bibinfo{booktitle}{\emph{Proceedings of the
  18th ACM SIGKDD international conference on Knowledge discovery and data
  mining}}. \bibinfo{pages}{786--794}.
\newblock


\bibitem[Lewis and Rao(2015)]%
        {lewis2015unfavorable}
\bibfield{author}{\bibinfo{person}{Randall~A Lewis} {and}
  \bibinfo{person}{Justin~M Rao}.} \bibinfo{year}{2015}\natexlab{}.
\newblock \showarticletitle{The unfavorable economics of measuring the returns
  to advertising}.
\newblock \bibinfo{journal}{\emph{The Quarterly Journal of Economics}}
  \bibinfo{volume}{130}, \bibinfo{number}{4} (\bibinfo{year}{2015}),
  \bibinfo{pages}{1941--1973}.
\newblock


\bibitem[Lewis et~al\mbox{.}(2011)]%
        {lewis2011here}
\bibfield{author}{\bibinfo{person}{Randall~A Lewis}, \bibinfo{person}{Justin~M
  Rao}, {and} \bibinfo{person}{David~H Reiley}.}
  \bibinfo{year}{2011}\natexlab{}.
\newblock \showarticletitle{Here, there, and everywhere: correlated online
  behaviors can lead to overestimates of the effects of advertising}. In
  \bibinfo{booktitle}{\emph{Proceedings of the 20th international conference on
  World wide web}}. \bibinfo{pages}{157--166}.
\newblock


\bibitem[Lewis and Reiley(2014)]%
        {lewis2014online}
\bibfield{author}{\bibinfo{person}{Randall~A Lewis} {and}
  \bibinfo{person}{David~H Reiley}.} \bibinfo{year}{2014}\natexlab{}.
\newblock \showarticletitle{Online ads and offline sales: measuring the effect
  of retail advertising via a controlled experiment on Yahoo!}
\newblock \bibinfo{journal}{\emph{Quantitative Marketing and Economics}}
  \bibinfo{volume}{12}, \bibinfo{number}{3} (\bibinfo{year}{2014}),
  \bibinfo{pages}{235--266}.
\newblock


\bibitem[Lewis and Wong(2018)]%
        {lewis2018incrementality}
\bibfield{author}{\bibinfo{person}{Randall~A Lewis} {and}
  \bibinfo{person}{Jeffrey Wong}.} \bibinfo{year}{2018}\natexlab{}.
\newblock \showarticletitle{Incrementality bidding \& attribution}.
\newblock \bibinfo{journal}{\emph{Available at SSRN 3129350}}
  (\bibinfo{year}{2018}).
\newblock


\bibitem[Liu et~al\mbox{.}(2020)]%
        {liu2020trustworthy}
\bibfield{author}{\bibinfo{person}{Min Liu}, \bibinfo{person}{Jialiang Mao},
  {and} \bibinfo{person}{Kang Kang}.} \bibinfo{year}{2020}\natexlab{}.
\newblock \showarticletitle{Trustworthy online marketplace experimentation with
  budget-split design}.
\newblock \bibinfo{journal}{\emph{arXiv preprint arXiv:2012.08724}}
  (\bibinfo{year}{2020}).
\newblock


\bibitem[Nath et~al\mbox{.}(2013)]%
        {nath2013ad}
\bibfield{author}{\bibinfo{person}{Abhirup Nath}, \bibinfo{person}{Shibnath
  Mukherjee}, \bibinfo{person}{Prateek Jain}, \bibinfo{person}{Navin Goyal},
  {and} \bibinfo{person}{Srivatsan Laxman}.} \bibinfo{year}{2013}\natexlab{}.
\newblock \showarticletitle{Ad impression forecasting for sponsored search}. In
  \bibinfo{booktitle}{\emph{Proceedings of the 22nd international conference on
  World Wide Web}}. \bibinfo{pages}{943--952}.
\newblock


\bibitem[Ockenfels et~al\mbox{.}(2006)]%
        {ockenfels2006online}
\bibfield{author}{\bibinfo{person}{Axel Ockenfels}, \bibinfo{person}{David~H
  Reiley~Jr}, {and} \bibinfo{person}{Abdolkarim Sadrieh}.}
  \bibinfo{year}{2006}\natexlab{}.
\newblock \bibinfo{title}{Online auctions}.
\newblock
\newblock


\bibitem[Parmar et~al\mbox{.}(2017)]%
        {parmar2017forecasting}
\bibfield{author}{\bibinfo{person}{Krunal Parmar}, \bibinfo{person}{Samuel
  Bushi}, \bibinfo{person}{Sourangshu Bhattacharya}, {and}
  \bibinfo{person}{Surender Kumar}.} \bibinfo{year}{2017}\natexlab{}.
\newblock \showarticletitle{Forecasting ad-impressions on online retail
  websites using non-homogeneous hawkes processes}. In
  \bibinfo{booktitle}{\emph{Proceedings of the 2017 ACM on Conference on
  Information and Knowledge Management}}. \bibinfo{pages}{1089--1098}.
\newblock


\bibitem[P{\"a}rssinen et~al\mbox{.}(2018)]%
        {parssinen2018blockchain}
\bibfield{author}{\bibinfo{person}{Matti P{\"a}rssinen}, \bibinfo{person}{Mikko
  Kotila}, \bibinfo{person}{Rub{\'e}n~Cuevas Rumin}, \bibinfo{person}{Amit
  Phansalkar}, {and} \bibinfo{person}{Jukka Manner}.}
  \bibinfo{year}{2018}\natexlab{}.
\newblock \showarticletitle{Is blockchain ready to revolutionize online
  advertising?}
\newblock \bibinfo{journal}{\emph{IEEE Access}}  \bibinfo{volume}{6}
  (\bibinfo{year}{2018}), \bibinfo{pages}{54884--54899}.
\newblock


\bibitem[Sahni(2015)]%
        {sahni2015effect}
\bibfield{author}{\bibinfo{person}{Navdeep~S Sahni}.}
  \bibinfo{year}{2015}\natexlab{}.
\newblock \showarticletitle{Effect of temporal spacing between advertising
  exposures: Evidence from online field experiments}.
\newblock \bibinfo{journal}{\emph{Quantitative Marketing and Economics}}
  \bibinfo{volume}{13}, \bibinfo{number}{3} (\bibinfo{year}{2015}),
  \bibinfo{pages}{203--247}.
\newblock


\bibitem[Tang et~al\mbox{.}(2010)]%
        {tang2010overlapping}
\bibfield{author}{\bibinfo{person}{Diane Tang}, \bibinfo{person}{Ashish
  Agarwal}, \bibinfo{person}{Deirdre O'Brien}, {and} \bibinfo{person}{Mike
  Meyer}.} \bibinfo{year}{2010}\natexlab{}.
\newblock \showarticletitle{Overlapping experiment infrastructure: More,
  better, faster experimentation}. In \bibinfo{booktitle}{\emph{Proceedings of
  the 16th ACM SIGKDD international conference on Knowledge discovery and data
  mining}}. \bibinfo{pages}{17--26}.
\newblock


\bibitem[Varian(2010)]%
        {varian2010computer}
\bibfield{author}{\bibinfo{person}{Hal~R Varian}.}
  \bibinfo{year}{2010}\natexlab{}.
\newblock \showarticletitle{Computer mediated transactions}.
\newblock \bibinfo{journal}{\emph{American Economic Review}}
  \bibinfo{volume}{100}, \bibinfo{number}{2} (\bibinfo{year}{2010}),
  \bibinfo{pages}{1--10}.
\newblock


\bibitem[Xu et~al\mbox{.}(2015)]%
        {xu2015smart}
\bibfield{author}{\bibinfo{person}{Jian Xu}, \bibinfo{person}{Kuang-chih Lee},
  \bibinfo{person}{Wentong Li}, \bibinfo{person}{Hang Qi}, {and}
  \bibinfo{person}{Quan Lu}.} \bibinfo{year}{2015}\natexlab{}.
\newblock \showarticletitle{Smart pacing for effective online ad campaign
  optimization}. In \bibinfo{booktitle}{\emph{Proceedings of the 21th ACM
  SIGKDD International Conference on Knowledge Discovery and Data Mining}}.
  \bibinfo{pages}{2217--2226}.
\newblock


\bibitem[Zhang et~al\mbox{.}(2014)]%
        {zhang2014optimal}
\bibfield{author}{\bibinfo{person}{Weinan Zhang}, \bibinfo{person}{Shuai Yuan},
  {and} \bibinfo{person}{Jun Wang}.} \bibinfo{year}{2014}\natexlab{}.
\newblock \showarticletitle{Optimal real-time bidding for display advertising}.
  In \bibinfo{booktitle}{\emph{Proceedings of the 20th ACM SIGKDD international
  conference on Knowledge discovery and data mining}}.
  \bibinfo{pages}{1077--1086}.
\newblock


\end{thebibliography}

\end{document}